\documentclass[prl,twocolumn,showpacs,superscriptaddress,preprintnumbers,amsmath,amssymb]{revtex4-1}

\usepackage{graphicx}
\usepackage{dcolumn}
\usepackage{bm}
\usepackage{hyperref}
\usepackage{amsmath}
\usepackage{amssymb}
\usepackage{subfigure}
\usepackage{epstopdf}
\usepackage{color}
\usepackage{wasysym}

\newcounter{defcounter}
\newenvironment{myequation}{%
\addtocounter{equation}{-1}
\refstepcounter{defcounter}

\begin{equation}}
{\end{equation}}

\begin{document}

\begin{abstract}
We consider an interacting, one-dimensional Bose gas confined in a split trap, obtained by an harmonic potential with a localized barrier at its center. We address its quantum-transport properties through the study of dipolar oscillations, which are induced by  a sudden quench of the position of the center of the trap. We find  that the dipole-mode  frequency strongly  depends on the interaction strength between the particles, yielding   information on the classical screening of the barrier and on its renormalization due to quantum fluctuations. Furthermore, we predict a parity effect  which becomes most prominent in the strongly correlated regime.
\end{abstract}

\pacs{05.30.Jp, 67.10.Jn, 67.85.-d, 03.75.Kk}

\title{Dipole mode of a strongly correlated one-dimensional Bose gas in a split trap: parity effect and barrier renormalization}

\author{Marco Cominotti}
\affiliation{Universit\'e Grenoble Alpes, LPMMC, F-38000 Grenoble, France}
\affiliation{CNRS, LPMMC, F-38000 Grenoble, France}

\author{Frank Hekking}
\affiliation{Universit\'e Grenoble Alpes, LPMMC, F-38000 Grenoble, France}
\affiliation{CNRS, LPMMC, F-38000 Grenoble, France}

\author{Anna Minguzzi}
\affiliation{Universit\'e Grenoble Alpes, LPMMC, F-38000 Grenoble, France}
\affiliation{CNRS, LPMMC, F-38000 Grenoble, France}

\maketitle

The study of elementary excitations is a fundamental aspect of many-body theories. For neutral quantum fluids, these excitations at  low energy correspond to sound waves for homogeneous systems and to inhomogeneous collective modes with discrete frequencies for confined ones. The analysis of the latter in ultracold quantum gases has been the subject of intense experimental~\cite{cornell,ketterle,cornell2,ketterle2,marago,inguscio,bismut,grimm} and theoretical~\cite{singh,stringari,stringari_scissors,minguzzi_fermions1,minguzzi_fermions2,fazio} activity in the last decades. A variety of different excitation modes has been characterized, as the monopole (breathing), dipole (sloshing), quadrupole, and scissor modes,  to cite the best known. An unprecedented precision has been reached in the measurement of their frequencies, becoming one of the most reliable tests for theoretical models and tools to investigate many-body phases and beyond-mean field effects~\cite{burnett,ketterle,altmeyer,salomon}. One of the most interesting aspects of these excitations is that  their frequencies depend on the microscopic properties of the system, yielding information e.g. on the equation of state or on its superfluid properties.

The analysis of collective modes allows in particular to  investigate the interplay between strong interactions and confinement in low-dimensional geometries~\cite{pitaevskii_rosch,menotti,pedri,olshanii_2D}.  In this work, we show that by using a specific confining geometry, i.e. a localized barrier at the center of a quasi-onedimensional harmonic trap,  one can directly access the effect of  quantum fluctuations, which play a major role in low dimensions. Effectively one-dimensional systems have been realized in ultracold-atoms experiments, by  employing optical lattices to create arrays of tubes or by creating a single atomic waveguide on an atom chip~\cite{esslinger,fertig,haller,fang}.   Strongly correlated phases are more accessible in one-dimensional gases~\cite{citro_RMP}, where interparticle interactions may be tuned by confinement-induced resonances, and because, counterintuitively, in one dimension the interactions become dominant in the low-density regime where particle losses and three-body recombination effects are reduced. The demonstration of the peculiar fermionized Tonks-Girardeau phase constitutes a beautiful example~\cite{paredes, kinoshita}.

We focus on the dipolar excitation mode of a one-dimensional (1D) ultracold Bose gas, i.e. on  a periodic oscillation of the center of mass of the atomic cloud.
In ultracold-gas experiments, this sloshing mode can be excited  by a displacement of  the center of the confining potential.  A localized barrier can be created by microscope-focused laser beams~\cite{desbusquois} or by a light-sheet repulsive potential~\cite{minardi,stroescu}. For a purely harmonic potential, as predicted by Kohn's theorem~\cite{kohn,dobson}, the dipole mode has the same frequency as the harmonic trap for arbitrary interactions.  In the presence of the barrier, Kohn's theorem does not apply. In this work we show that the dipole mode displays an interaction-dependent frequency shift which allows to estimate directly the effective barrier strength seen by the fluid.  We also find a surprising parity effect in the oscillation frequency, which  becomes important in the strongly correlated phase and can be understood in terms of fermionic rather than bosonic transport processes. The dipolar oscillation of the cloud 
 realizes in fact a specific type of quantum transport across the barrier.  Quantum transport phenomena are more and more explored  with ultracold atoms~\cite{esslinger_transport2,minardi_fazio,esslinger_transport1}.

\begin{figure}[b]
  \includegraphics[width=0.49\textwidth]{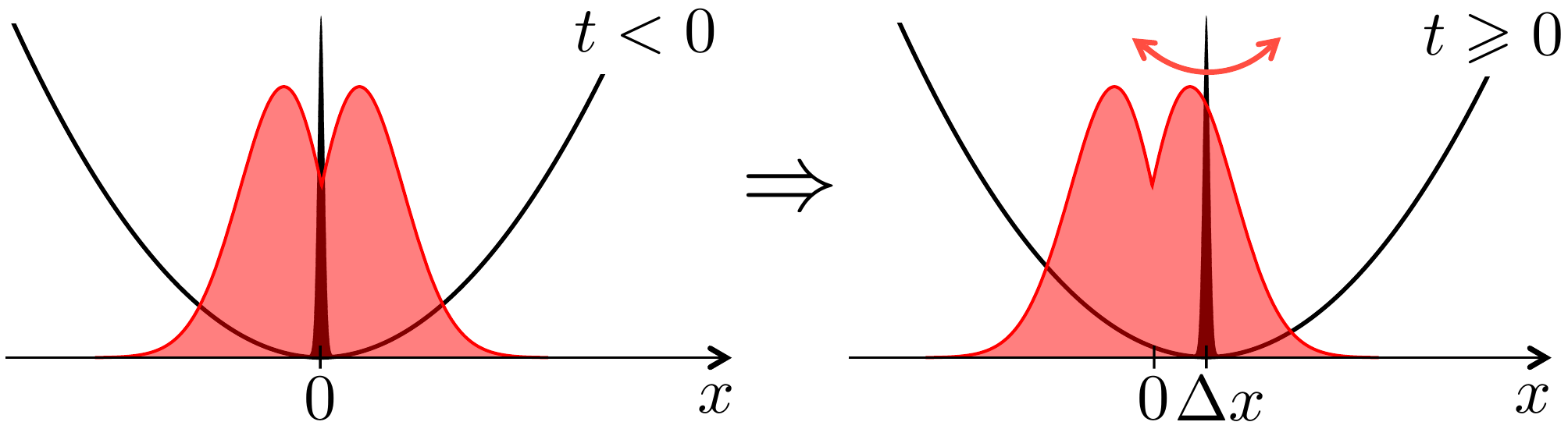}
  \caption{(Color online). Sketch of the potential along the 1D wire. At $t=0$ the potential is displaced by $\Delta x$ to induce the dipole oscillation of the center of mass of the Bose gas. }
  \label{cartoon}
\end{figure}

{\em System and Main Results}.--- We consider
a tight atomic waveguide, containing $N$ bosons
of mass $m$ at zero temperature, confined to a 1D geometry. The bosons interact with each other via a contact
interaction  $v(x-x')=g \, \delta(x-x')$, and are subjected to the harmonic confinement $V_{\rm h}(x)=\frac{1}{2}m\omega_{\rm h}^{2}x^2$ along the waveguide. The waveguide also contains a localized potential barrier at the center of the harmonic confinement, $V_{\rm b}(x)=U_0\delta(x)$, which gives rise to a split trap for the particles. 
The corresponding Hamiltonian reads
\begin{equation}
\mathcal{H}=\sum_{j=1}^{N} -\frac{\hbar^2}{2m}
  \frac{\partial^2}{\partial^{2} x_{j}}
  + U_0\delta(x_j) + \frac{1}{2}m\omega_{\rm h}^{2}x_{j}^{2}+ \frac{g}{2} \! \sum_{j,l=1}^{N}\delta(x_{l}-x_{j}) \,.
  \label{eq:hamiltonian}
\end{equation}

\begin{figure}[t]
  \includegraphics[width=0.40\textwidth]{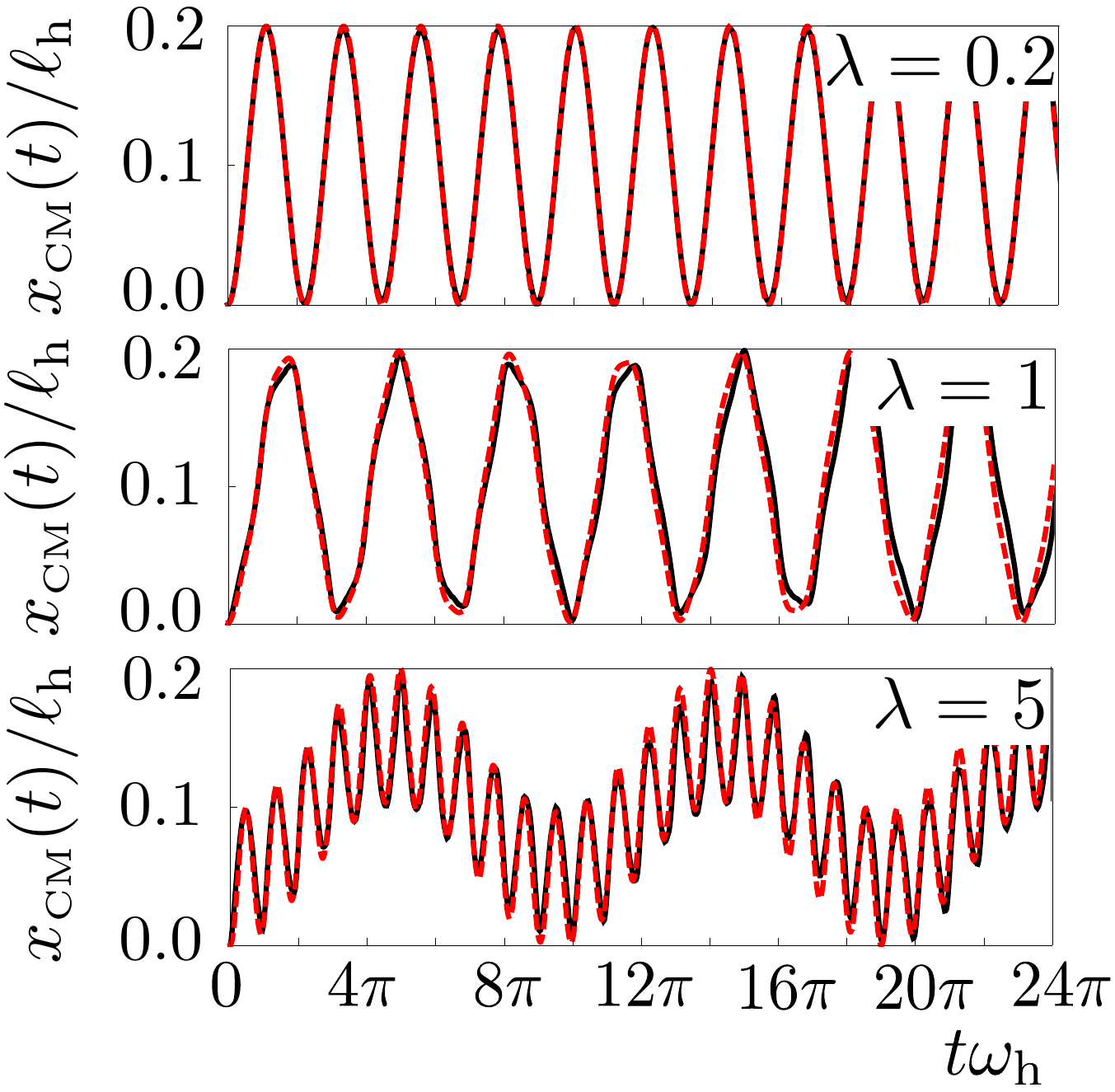}
  \caption{(Color online). Time evolution of the center of mass position $x_{\text{\tiny{CM}}}(t)$ for various values of barrier strength $\lambda=U_0/E_{\rm h}\ell_{\rm h}$, with $E_{\rm h}=\hbar\omega_{\rm h}$,   and for interaction strength $g=0$. Solid black line: real-time evolution obtained from the numerical solution of the quench problem, red dashed line: estimate using the energy gaps of the many-body spectrum -- a single frequency $(E_{1}-E_{0})/\hbar$ in the first panel, and an additional frequency for the second and third panel.}
  \label{oscill}
\end{figure}

As sketched  in Fig.~\ref{cartoon}, starting from the ground state in the split trap,  we study the dynamics following a sudden quench of the harmonic and barrier potential positions, both being displaced at time $t=0^+$ by a small amount $\Delta x \ll \ell_{\rm h}$, with $\ell_{\rm h}=\sqrt{\hbar/m\omega_{\rm h}}$ the characteristic harmonic oscillator length. This induces a collective oscillation of the gas of dipolar nature, and in particular the center of mass (CM) of the gas starts to oscillate periodically in time with a main frequency $\omega_{\rm d}$ around the new center of the trap. For specific interaction regimes (non-interacting, weakly, and infinitely strongly interacting) we have found the full time-resolved dynamical evolution of the cloud following the quench, an example being illustrated in Fig.~\ref{oscill}.

In the absence of the barrier, when only the harmonic confinement is present, the oscillation  is purely sinusoidal: only the sloshing mode is excited by the quench, at the frequency of the harmonic confinement. This Kohn's theorem result~\cite{kohn,dobson}, specific to the harmonic confinement, is valid for arbitrary interaction between the particles, which can be understood as a change of reference frame into an accelerated one \cite{minguzzi2001}.

\begin{figure*}[t]
  \includegraphics[width=0.98\textwidth]{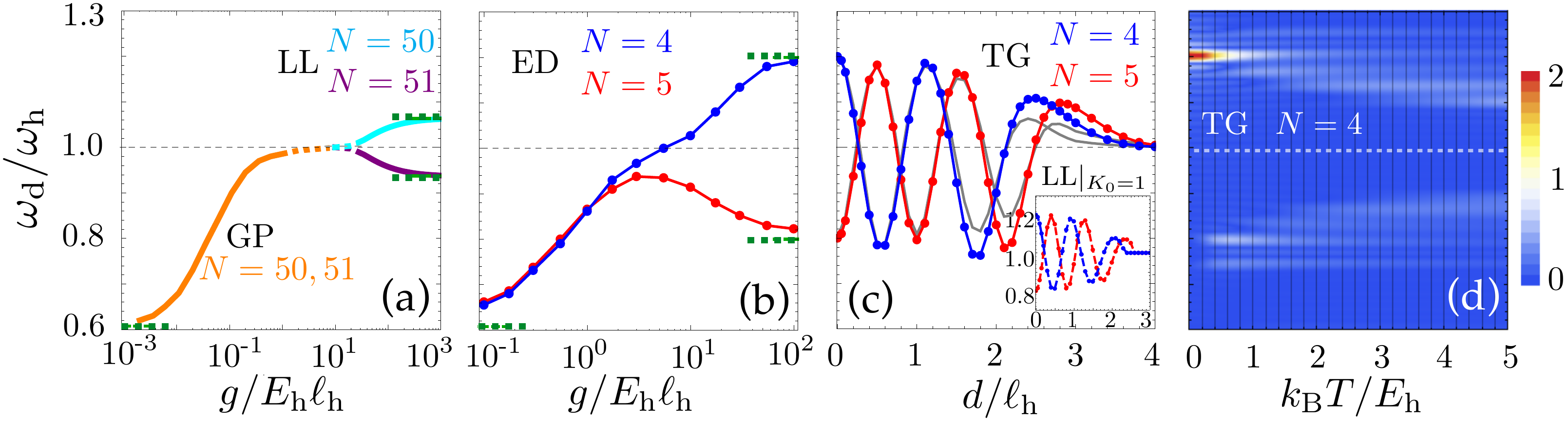}
  \caption{(Color online). (a, b) Frequency of the dipole mode $\omega_{\rm d}$ as function of the interaction strength $g$ for $U_0/E_{\rm h}\ell_{\rm h}=1$, in different regimes: (a) GP and LL solution for $N=50, 51$ (orange, cyan and purple lines respectively), the dotted lines at intermediate interactions are a guide to the eye in the regime beyond the validity of the GP and LL approaches;  (b) ED solution for $N=4, 5$ (blue and red respectively). The green dashed lines in the panels (a, b) correspond to the limiting cases of NI and TG gas ($g\rightarrow 0, \infty$). (c) Dependence of $\omega_{\rm d}$ on the position of the barrier $d$, for $g\rightarrow \infty$; calculated with the exact TG technique and corresponding LL for $K_{0}=1$ (inset, with the LDA density, and gray lines, using the exact TG density). (d) Density plot of the Fourier transform of the center of mass oscillation $|\Re\mbox{e}[\delta x_{\text{\tiny{CM}}}(\omega)]|$ as a function of frequency and temperature, in the TG limit, for $U_0/E_{\rm h}\ell_{\rm h}=1$ and $N=4$.}
  \label{omega}
\end{figure*}

In the presence of the barrier  (see again Fig.~\ref{oscill}), we observe the appearance of an additional harmonic component and a frequency shift of the dipolar oscillation with respect to the harmonic confining one. This frequency shift, which is the main topic of our analysis, increases with the barrier strength, and  depends on the interaction regimes between the particles (see  Fig.~\ref{omega}(a, b)). The frequency shift directly reflects the strength of the renormalized barrier seen by the fluid. The barrier is maximally reduced at intermediate interactions as a consequence of the competition of classical screening -- occurring for weak interactions and {\em increasing} with the interaction strength, and barrier renormalization by the quantum fluctuations of the density -- occurring for strong interactions and, being conjugate to phase fluctuations, {\em decreasing} for increasing interactions. Furthermore, we find that the frequency shift of the dipole mode in the strongly correlated regime depends on the particle number being even or odd.  This is particularly striking for a bosonic system, which does not display parity effects in other observables e.g. in the persistent currents \cite{manninen2,cominotti}. When the barrier potential is not placed at the center of the harmonic trap, the parity effect is still present, but it is modulated by the position of the barrier, see Fig.~\ref{omega}(c) and~\cite{supp}. Remarkably, we find also that signatures of this mesoscopic effect remains visible at finite temperatures, Fig.~\ref{omega}(d).

We detail below the steps and methods employed to obtain Figs.~\ref{oscill} and~\ref{omega}.

{\em Exact solutions}.--- In the two limiting cases of noninteracting (NI) and Tonks-Girardeau (TG) gas, we find an exact solution for the dynamical evolution of the gas.
In the TG case, the solution is obtained by mapping the system onto a gas of non-interacting fermions subjected to the same external (time-dependent) potential ~\cite{girardeau,girardeauTD}. In both cases, starting from the analytical expression for the ground-state wavefunction $\Psi_{0}^{t< 0}$ before the quench, we  perform a numerical evolution of the ground-state wave function in real time $|\Psi_{0}^{t\geqslant 0}(t)\rangle=\exp(-i\mathcal{H}^{t\geqslant 0}t/\hbar)|\Psi_{0}^{t< 0}\rangle$. The resulting time evolution of the position of the CM $x_{\text{\tiny{CM}}}(t)=\int {\rm d}x \, x \,n(x,t)$, where $n(x,t)$ is the time-dependent gas density, is illustrated in Fig.~\ref{oscill}.  A Fourier analysis of $x_{\text{\tiny{CM}}}(t)$ shows that the center of mass motion is characterized by a main dipolar frequency, corresponding to the difference between the energy of the first excited many-body state and of the ground state of (\ref{eq:hamiltonian}),  $\omega_{\rm d}=(E_{1}-E_{0})/\hbar$, as well as by a higher harmonic with frequency $(E_{3}-E_{0})/\hbar$, in agreement with many-body perturbation theory \cite{supp}. 

In the following we shall focus on the dipolar frequency $\omega_{\rm d}$ and evaluate it in various interaction regimes.
For the NI case, the $N$-particle ground-state energy is given by
$E^{\rm NI}_{0}= N \varepsilon_0$, and the first excited one by $E^{\rm NI}_{1}= (N-1) \varepsilon_{0}+\varepsilon_{1}$, where $\varepsilon_j$ are the single particle energies, solutions of 
\begin{equation} \left[-\frac{\hbar^2}{2m}\partial^{2}_{x}
+ U_0\delta(x) + \frac{1}{2}m\omega_{\rm h}^{2}x^{2}\right] \psi_n = \varepsilon_n \psi_n\;,
\label{oneschrod}
\end{equation}
which can be expressed in terms of Hermite and Wittaker functions~\cite{supp,bush}. In the TG case, the many-body energy spectrum coincides with the one of a Fermi gas. In particular, the ground state energy is $E^{\rm TG}_{0}=\sum_{k=0}^{N-1} \varepsilon_k$ and the first excited state one is $E^{\rm TG}_{1}=\sum_{k=0}^{N-2} \varepsilon_k+\varepsilon_N$. Both NI and TG limits are indicated in Fig.~\ref{omega}(a, b).

The TG solution allows to readily get  physical insight on the parity effect. Using perturbation theory in the barrier strength we have $\hbar\omega_{\rm d}=E^{\rm TG}_{1}-E^{\rm TG}_{0}= \hbar \omega_{\rm h} +\langle \Psi^{\rm TG}_{1}|\mathcal{H}_{\rm b}|\Psi^{\rm TG}_{1}\rangle-\langle \Psi^{\rm TG}_{0}|\mathcal{H}_{\rm b}|\Psi^{\rm TG}_{0}\rangle$, where $\mathcal{H}_{\rm b}=\sum_{j=1}^{N}U_0\delta(x_{j})$. Using the explicit form of the ground- and first-excited many body wavefunction in the TG limit, $\Psi^{\rm TG}(x_1,...x_N)=\Pi_{1\le j<k\le N}{\rm sign}(x_j-x_k) \det[\psi_n(x_\ell)]$, where $n=0...N-1$ for the ground state and $n=0...N-2,N$ for the first excited state, we readily obtain 
\begin{equation}
\hbar\omega_{\rm d}=\hbar\omega_{\rm h}+U_0(|\psi_{N}(0)|^{2}-|\psi_{N-1}(0)|^2)\,. 
\end{equation}
The single-particle orbitals $\psi_{k}(0)$ vanish for $k$ odd and are finite for $k$ even~\cite{supp}, therefore $\omega_{\rm d}$ is larger or smaller than $\omega_{\rm h}$ depending on the number of particles being even or odd. In analogy to transport phenomena in Fermi gases,  for the strongly correlated (fermionized) bosons  the dynamics is determined by the states at the (effective) Fermi level. By using the analytical expression of $\psi_{k}$ we also find  that in the TG regime the frequency shift $|\omega_{\rm d}-\omega_{\rm h}|$ scales as $1/\sqrt{N}$ for large $N$ (see~\cite{supp} for details as well as for the nonperturbative calculation in the TG regime).  As we shall see below, this coincides with a special case of the LL prediction.
If the barrier is placed at a distance $d$ from the center of the harmonic trap, the TG solution allows  to calculate exactly the dipole frequency by solving the modified Eq.~(\ref{oneschrod}) with barrier potential $U_0\delta(x-d)$~\cite{supp,bush2}. As shown in Fig.~\ref{omega}(c) the parity effect is modulated, displaying an oscillating behavior as a function of $d$.

{\em Exact diagonalization} (ED).--- To cover all the interaction strength regimes,   for small numbers of particles, we use a numerical method based on the exact diagonalization of the Hamiltonian (\ref{eq:hamiltonian}): We calculate the low-energy eigenspectrum of the many-body system, and  obtain the dipole mode frequency as $\omega_{\rm d}=(E_{1}-E_{0})/\hbar$.
To represent the  Hamiltonian~(\ref{eq:hamiltonian}) we have chosen the $N$-particle basis built using the single-particle eigenfunctions of Eq.~(\ref{oneschrod}), which are analytically known. The truncation of the Hilbert space  to a number $S$ of single-particle states  is the only approximation performed with this technique. Since the dimension of the Hilbert space rapidly grows with $S$  and the particle number $N$  according to  $\binom{S+N-1}{N}$, we are limited to small $N$~\footnote{We take systems up to $N=5$ and $S=18$, corresponding to a size of the Hilbert space of $26334$. To improve the estimate of the eigenspectrum obtained with the truncated Hilbert space we perform a finite-size scaling.}. The predictions of ED calculations are shown in Fig.~\ref{omega}(b).

{\em Mean field approach}.--- For larger values of the particle number we adopt complementary approximate approaches. In the regime of weak interactions, neglecting quantum fluctuations, we describe the fluid as a Bose-Einstein condensate~\cite{petrov} by means of the mean-field Gross-Pitaevskii (GP) equation:
$ [-\frac{\hbar^2}{2m} \partial^{2}_{x}  + U_0\delta(x) + \frac{1}{2}m\omega_{0}^{2}x^{2}+g|\Phi|^2 ]\Phi = \mu\Phi$,
where $\Phi(x)$ is the condensate wave function, normalized to the particle number $N$, and $\mu$ the chemical potential.
We integrate this equation in imaginary time to find its ground state solution, and, subsequently, we evolve the ground state in real time with the shifted Hamiltonian, to calculate the time evolution of the position of the CM, and thus the frequency of the dipole mode in Fig.~\ref{omega}(a).

{\em Luttinger liquid theory}.--- In the strongly interacting regime  we take into account the effect of quantum
fluctuations using the Luttinger liquid (LL) theory~\cite{haldane}. This is a
a low-energy, quantum hydrodynamics description of the bosonic fluid,
in terms of the canonically conjugate fields $\theta$ and $\phi$, corresponding to fluctuations of the density and phase, respectively.
The presence of the smoothly varying harmonic potential is  taken into account  within the local density approximation (LDA), i.e. expressing the equation of state of the inhomogeneous system as a functional of the local density $n(x)$.
The effective LL Hamiltonian for the inhomogeneous system reads~\cite{cazalilla,citro}
\begin{equation}
  \mathcal{H}^{\rm{LL}}_{0}\!=\!\!\frac{\hbar}{2\pi}\!\!\int_{-\infty}^{\infty}\!\!\!\!\!\mbox{d}x\!\left[v_{s}(x)K(x)(\partial_{x}\phi(x))^2\!+\!\frac{v_{s}(x)}{K(x)}(\partial_{x}\theta(x))^2 \right]\!.
  \label{eq:llhamiltonian}
\end{equation}
The local Luttinger parameter $K(x)$ and the sound velocity $v_{s}(x)$ depend on the microscopic interaction strength $g$, and are determined  according to $v_{s}(x)K(x)=\hbar\pi n(x)/m$, and $v_{s}(x)/K(x)=\partial_{n}\mu(n(x))/\hbar\pi$. 
We assume an equation of state of the form $\mu(n)=\eta n^{\nu}$ \cite{menotti,citro}: this connects the GP regime, where $\mu(n)=gn$, and TG regime, where $\mu(n)=(\hbar^{2}\pi^{2}/2m)n^2$. The parameters $\eta$ and $\nu$ are obtained from the  Bethe ansatz solution of the homogeneous Lieb-Liniger model~\cite{dunjko,supp}. For this model, the dependence on the interaction strength of the Luttinger parameters is known (see, e.g.~\cite{cazalilla}); let us recall in particular that in the TG limit $K_0=1$ and $v_s=v_F$, the Fermi velocity of the fermionized Bose gas. 

The localized barrier potential induces a very rapid spatial variation of the confinement, hence cannot be taken into account in the LDA description. We treat its effect  perturbatively in the limit of weak barrier strength. The barrier, located in full generality at position $d$, yields a non-harmonic contribution to the Luttinger liquid Hamiltonian of the form $\mathcal{H}^{\rm LL}_{\rm{b}}=\int_{-\infty}^{\infty} \mbox{d}x\, U_0 \delta(x-d) \rho(x)$. Keeping only the lowest harmonics in the density field expansion
$\rho(x)=[n(x)+\partial_{x}\theta(x)/\pi]\sum_{l=-\infty}^{+\infty}e^{2il\theta(x)+2il\pi \int_{-\infty}^{x}{\rm d} x'n(x')}$, we obtain the dominant term $\mathcal{H}^{\rm LL}_{\rm{b}}\sim 2 U_{0} n(d)\cos[2\theta(d)+2\pi \int_{-\infty}^{d}\mbox{d}x\, n(x)]$.
Integrating out the high-energy modes of the field $\theta(x)$, we obtain a renormalization of the barrier strength by quantum fluctuations, such that the effective barrier is given by $U^{\rm eff}=U_{0}e^{-2G(d)}$, where $G(x)$ is the local two point correlation function for $\theta(x)$. In the particular case  where the barrier is at the center of the trap and for large particle numbers we find  $U^{\rm eff}=U_{0}\left(\frac{a}{N}\right)^\kappa$, where $\kappa=K_{0}\sqrt{\nu/2}$, and $a$ is a non-universal parameter dependent on the cut-off of the effective LL theory~\footnote{We have used as estimate for $a$ its value in the TG limit, accessible through the exact solution.}. Perturbation theory~\cite{supp} finally yields
\begin{equation}
\label{eq:wLL}
\omega_{\rm d}= \omega_{{\rm h}} +\frac{n(d)U^{\rm{eff}}(d)K(d)v_{s}(d)}{\hbar \omega_{\rm h}R}f(\nu)\cos\!\left(\!2\pi \mbox{$\int_{-\infty}^{d}$}\mbox{d}x\, n(x)\!\right)\!,
\end{equation}
where $f(\nu)=(\frac{1}{\nu}\!+\!\frac{1}{2})\Gamma^{2}(\frac{1}{\nu}\!+\!\frac{1}{2})2^{\frac{2}{\nu}+2}/\Gamma(\frac{2}{\nu}\!+\!1)$.
 The integral of the density gives rise to an oscillating term as a function of $d$ shown  in Fig.~\ref{omega}(c), which reduces to  $(-1)^{N}$  for $d=0$. This explains the  parity-dependent frequency shift of the dipole mode. 
We have used Eq.(\ref{eq:wLL}) to estimate the interaction-dependent frequency shift of the dipole mode in Fig.~\ref{omega}(a).  For repulsive interactions $\kappa\geqslant 1$, and the effective barrier strength is smaller than the bare one, decreasing as the interaction strength is decreased from infinite to intermediate values ($\kappa\rightarrow 1$ for $g\rightarrow\infty$, and $\kappa\rightarrow \infty$ for $g\rightarrow 0$), which explains the approaching of $\omega_{\rm d}$ to $\omega_{\rm h}$ at decreasing interactions. Finally, we notice that in the TG limit one has  $\kappa=1$ and $n^{\rm{TG}}(0)=(1/\ell_{\rm h})\sqrt{2N/\pi}$ (in LDA), thereby recovering the $|\omega_{\rm d}-\omega_{\rm h}|\propto 1/\sqrt{N}$ scaling behavior found with the TG exact solution.

{\em  Experimental considerations}.--- The above analysis is readily extended to include both thermal effects and a finite width of the barrier. We have performed an exact finite-temperature calculation in the TG limit~\cite{supp}, see Fig.~\ref{omega}(d). Interestingly, the main features of the dipole-mode frequency shift remain visible at finite temperatures: for temperatures $k_{\rm B}T \gtrsim \hbar \omega_{\rm h}$ thermal fluctuations mix the characteristic zero-temperature frequencies of odd and even number of particles. Finally, we have also checked that a finite barrier width of the order of the interparticle spacing does not considerably affect the estimate of the frequency shift~\footnote{Considering a Gaussian barrier of width $\sigma/\ell_{\rm h}\leqslant 0.2$, we have found corrections to the dipole frequency below $5\%$.}.

{\em Conclusions.}--- As a prototype of quantum transport in an interacting 1D Bose gas, we have studied the dipolar oscillations of the gas across a localized barrier, induced by a sudden shift of the trap center. We have found that the main frequency of the oscillation allows to determine the effective barrier strength seen by the fluid. The full quantum solution also displays a peculiar parity effect, due to the combination of fermionic transport properties for the correlated Bose gas and the harmonic trap geometry. Observation of this parity effect and of the shift of the dipole frequency  with the interaction strength would provide a non-ambiguous evidence of the effect of quantum fluctuations.

{\em Acknowledgments.} We wish to thank D. Basko, F. Minardi, M. Rizzi and D. Rossini for very stimulating discussions. M.C. is indebted to F. Calcavecchia for help with the numerics. This work is supported by the ERC Handy-Q grant N.258608, and by the Institut Universitaire de France.


\newpage

\appendix

\onecolumngrid

\begin{center}
{\large \textbf{SUPPLEMENTAL MATERIAL FOR: \\
``Dipole mode of a strongly correlated one-dimensional Bose gas in a split trap: parity effect and barrier renormalization"}}
\end{center}

\section{I. Dipole mode frequency from many-body perturbation theory}
\label{sec1}
We consider an excitation of the center of mass (CM) of the system   via a sudden displacement of the external harmonic and barrier potentials $V_{\rm ext}=\sum_{j=1}^{N}\frac{1}{2}m\omega^{2}_{\rm h}x^{2}_{j}+U_{0}\delta{(x_{j})}$ by an  amount $\Delta x$ at time $t=0$.  For $t<0$ the system is assumed to be in its many-body ground state  $|\Psi_{0}^{t< 0}\rangle$ of  Hamiltonian   $\mathcal{H}^{t< 0}$ before the quench.  In accordance to perturbation theory, we decompose the initial ground state into the eigenstates of the system after the quench, $|\Psi_{0}^{t< 0}\rangle=|\Psi_{0}^{t\geqslant 0}\rangle+\sum_{k=1}^{\infty} c_{k} |\Psi_{k}^{t\geqslant 0}\rangle$, where $\mathcal{H}^{t\geqslant 0} |\Psi_{k}^{t\geqslant 0}\rangle= E_{k}^{t\geqslant 0} |\Psi_{k}^{t\geqslant 0}\rangle$ and $c_{k}= \langle\Psi_{k}^{t\geqslant 0}| \mathcal{H}^{t< 0}-  \mathcal{H}^{t\geqslant 0} |\Psi_{0}^{t\geqslant 0} \rangle/(E_{0}^{t\geqslant 0}-E_{k}^{t\geqslant 0})$. Assuming the displacement small, $\Delta x \ll\ell_{\rm h}$, we have $\mathcal{H}^{t< 0}\simeq \mathcal{H}^{t\geqslant 0}+\Delta x \sum_{j=1}^{N}\partial_{x_{j}}V_{\rm ext}^{t\geqslant 0}(x_{j})+ {\cal O}(\Delta x^2)$, where $\partial_{x_{j}}V_{\rm ext}^{t\geqslant 0}(x_{j})=m\omega_{\rm h}^{2}x_{j}+U_0\delta'(x_{j})$. The expectation value of the position of the center of mass  as a function of time is given by  $x_{\text{\tiny{CM}}}(t)=\int \mbox{d}x\,x \, n(x,t)$, in terms of the density  $n(x,t)=\int\mbox{d}x_{1},\dots,\mbox{d} x_{N} \sum_{j=1}^{N}\delta(x-x_{j})|\Psi_{0}^{t\geqslant 0}(x_{1},\dots,x_{N},t)|^{2}$. Perturbation theory finally yields
\begin{myequation}
x_{\text{\tiny{CM}}}(t)=x_0 + \int \mbox{d}x\,x \, \int\mbox{d}x_{1},\dots,\mbox{d} x_{N} \sum_{j=1}^{N}\delta(x-x_{j}) \sum_{k=1}^{\infty}c_{k} \,2 \, \Re\mbox{e}[\Psi^{*}_{0}(x_{1},\dots,x_{N})\Psi_{k}(x_{1},\dots,x_{N})e^{-i(E_{k}^{t\geqslant 0}-E_{0}^{t\geqslant 0})t/\hbar}].
\label{eq:cm1}
\end{myequation}
From this expression we see that the periodic oscillation of the center of mass motion is decomposed in a Fourier series at frequencies $(E_{k}^{t\geqslant 0}-E_{0}^{t\geqslant 0})/\hbar$, where the weight of each component depends on the coefficient $c_k$ and the overlap integrals of the many-body wavefunctions $\Psi^{*}_{0}\Psi_{k}$.

In the absence of the barrier, i.e. $U_0=0$, it follows from  Kohn's theorem that the only non vanishing matrix element $c_k$ is the $k=1$ one, as we have checked in our numerical solution. Therefore the center of mass evolves at a single frequency $\omega_{\rm d}=(E^{t\geqslant 0}_{1}-E^{t\geqslant 0}_{0})/\hbar$, ie the dipole one, where in particular we have that  $\omega_{\rm d}= \omega_{\rm h}$. 

In the presence of the barrier, i.e.  $U_0>0$, the  matrix elements $c_k$ for $k>1$ are in general  non vanishing and  the dynamics of the center of mass  will be determined by more frequencies $(E^{t\geqslant 0}_{k}-E^{t\geqslant 0}_{0})/\hbar$.  However, for small $U_0$, the contribution of the higher energy states with $k\geqslant 1$, gets less and less important, the most important contribution being the dipolar one. The dependence of the matrix element $c_{k}$ on the quantum number $k$ is shown in Fig.~\ref{fig:ck0} below. The overlap integrals in Eq.(\ref{eq:cm1}) further decrease the weight of the higher-frequency components.

 In conclusion, even in the presence of the barrier, we obtain that the sudden quench of the center of the trap excites mainly the dipole mode. The second harmonics of the motion is also obtainable with this method and well agree with Fig. 2 of the main text.

\begin{figure*}[h]
  \includegraphics[width=0.98\textwidth]{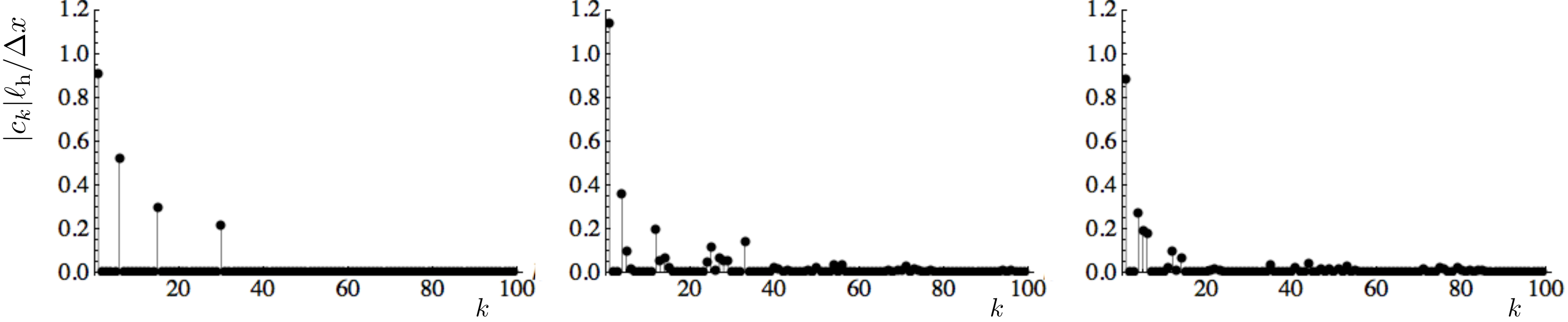}
  \caption{Absolute value of the matrix element $c_{k}$ as a function of the quantum number $k$, for the same value of dimensionless barrier strength considered in the main text $\lambda=1$, and different interparticle interaction strengths $g/E_{\rm h}\ell_{\rm h}=0, 3, 30$ (from left to right).}
  \label{fig:ck0}
\end{figure*}

\section{II. Details on the exact solution of the single-particle problem}
\label{exactsol}
\subsection{A. Exact solution for a centred barrier}
The one-body Schr\"odinger equation in the presence of harmonic and centred barrier potentials reads
\begin{myequation} 
 \left[-\frac{\hbar^2}{2m}\partial^{2}_{x}
+ U_{0}\delta(x) + \frac{1}{2}m\omega_{\rm h}^{2}x^{2}\right] \psi_n = \varepsilon_n \psi_n\;.
\label{oneschrod}
\end{myequation}
Scaling all the quantities in units of the harmonic oscillator level spacing $E_{\rm h}=\hbar\omega_{\rm h}$ and characteristic length $\ell_{\rm h}=\sqrt{\hbar/m\omega_{\rm h}}$ we obtain the Schr\"odinger equation in the reduced form
\begin{myequation}
\left[-\frac{1}{2}\partial^{2}_{x}
+ \lambda\delta(x) + \frac{1}{2}x^{2}\right] \psi_n = \varepsilon_n \psi_n\;,
\label{oneschrod_reduced}
\end{myequation}
where $\lambda=U_{0}/E_{\rm h}\ell_{\rm h}$.
The odd solutions $\psi_{2n+1}$ of the harmonic oscillator without barrier ($\lambda=0$) are still solution of this equation with barrier, and having a node at the position of the barrier, they are independent of the strength of the barrier potential.
The even solutions $\psi_{2n}$ can be expressed in terms of Whittaker functions, after noticing that Eq.~(\ref{oneschrod_reduced}) for $x>0$ corresponds to the differential equation for the parabolic cylinder functions~\cite{bush, abramowitz}:
\begin{myequation}
\begin{split}\psi_{2n}(x)&=\mathcal{N}D_{\varepsilon_{n}}(|x|)\;,\\
D_{\varepsilon_{n}}(x)&=\cos\left(\frac{\pi}{4}+\frac{\pi}{2}\varepsilon_{n}\right)Y_{1}-\sin\left(\frac{\pi}{4}+\frac{\pi}{2}\varepsilon_{n}\right)Y_{2}\;,	
\end{split}
\label{cylinderfunc}
\end{myequation}
where $\mathcal{N}$ is a normalization constant, and
\begin{eqnarray}
\nonumber Y_{1}&=&\frac{\Gamma\left(\frac{1}{4}-\frac{1}{2}\varepsilon_{n}\right)}{\sqrt{\pi}2^{\frac{1}{2}\varepsilon_{n}+\frac{1}{4}}}e^{-\frac{x^2}{2}}M\left(\frac{1}{2}\varepsilon_{n}+\frac{1}{4},\frac{1}{2},x^{2}\right)\;,\\
\nonumber Y_{2}&=&\frac{\Gamma\left(\frac{3}{4}-\frac{1}{2}\varepsilon_{n}\right)}{\sqrt{\pi}2^{\frac{1}{2}\varepsilon_{n}-\frac{1}{4}}}e^{-\frac{x^2}{2}}\sqrt{2}xM\left(\frac{1}{2}\varepsilon_{n}+\frac{3}{4},\frac{3}{2},x^{2}\right)\;.
\end{eqnarray}
Here, $M(a,b,c)$ is the confluent hypergeometric function and $\Gamma(a)$ is the complete $\Gamma$-function.\\
Imposing the cusp condition at the position of the barrier, $\partial_{x}\psi_{n}(0^{+})-\partial_{x}\psi_{n}(0^{-})=\lambda\psi_{n}(0)$, we get the following expression for the eigenvalues $\varepsilon_n$:
\begin{myequation}
 \frac{\Gamma\left( \frac{3}{4}-\frac{1}{2}\varepsilon_{n}\right)}{\Gamma\left( \frac{1}{4}-\frac{1}{2}\varepsilon_{n}\right)}=-\frac{\lambda}{2}\;.
 \end{myequation} 
We observe that for $\lambda\rightarrow 0$ the energy eigenvalues $\varepsilon_{n}$ tend to the even eigenvalues of the harmonic oscillator without barrier. In the opposite limit $\lambda\rightarrow \infty$ they tend to the odd eigenvalues, that become doubly degenerate.

\subsection{B. Exact solution for a non-centred barrier}

The one-body Schr\"odinger equation in the presence of harmonic and non-centred barrier potentials reads
\begin{myequation} 
 \left[-\frac{\hbar^2}{2m}\partial^{2}_{x}
+ U_{0}\delta(x-d) + \frac{1}{2}m\omega_{\rm h}^{2}x^{2}\right] \psi_n = \varepsilon_n \psi_n\;.
\label{oneschrodd}
\end{myequation}
In this case the parity symmetry of the system is broken, and we can not distinguish between even and odd solutions. The solution can be written piecewise in the form:
\begin{myequation} \psi_{n}(x)=\mathcal{N}_{l}\Theta(d-x)D_{\varepsilon_{n}}(-x)+\mathcal{N}_{r}\Theta(x-d)D_{\varepsilon_{n}}(x)\;,\end{myequation}
where $\Theta(x)$ is the Heaviside function, and $D_{\varepsilon_{n}}(x)$ are the parabolic cylinder functions defined in Eq.~(\ref{cylinderfunc}).
The condition of continuity and the cusp condition
at the position of the barrier lead to the trascendental equation for the energy eigenvalues~\cite{bush2}:
\begin{myequation} \left(\varepsilon_{n}-1/2\right)(D_{\varepsilon_{n}-1}(-d)D_{\varepsilon_{n}}(d)+D_{\varepsilon_{n}-1}(d)D_{\varepsilon_{n}}(-d))=\sqrt{2}\lambda D_{\varepsilon_{n}}(-d)D_{\varepsilon_{n}}(d)\;.\end{myequation}

\section{III. Details on the inhomogeneous Luttinger liquid description}
In order to diagonalize the inhomogeneous Luttinger liquid (LL) Hamiltonian in the presence of harmonic confinement, Eq. (4) of the main text, we have introduced, in analogy to what is done for the quantum harmonic oscillator, the bosonic conjugate fields $b_{j}$ and $b^{\dagger}_{j}$ in the following way:
\begin{myequation}
\begin{split}
\partial_{x}\phi(x,t)&=\sum_{j=0}^{\infty}i\sqrt{\frac{m\omega_{j}}{2\hbar n(x)}}\left(\varphi_{j}(x)e^{i\omega_{j}t}b^{\dagger}_{j} -\varphi^{*}_{j}(x)e^{-i\omega_{j}t}b_{j}\right)\;,\\
-\frac{\theta(x,t)}{\pi}&=\sum_{j=0}^{\infty}\sqrt{\frac{\hbar n(x)}{2m\omega_{j}}}\left(\varphi_{j}(x)e^{i\omega_{j}t}b^{\dagger}_{j} +\varphi^{*}_{j}(x)e^{-i\omega_{j}t}b_{j}\right)\;;
\label{modes}
\end{split}
\end{myequation}
such that the canonical commutation relation $[-\theta(x)/\pi,\partial_{x'}\phi(x')]=i\delta(x-x')$ and the orthogonality relation $\int\mbox{d}x \varphi^{*}_{j}(x)\varphi_{l}(x)=\delta_{j,l}$ imply that $[b_{j},b^{\dagger}_{l}]=\delta_{j,l}$.
Hamiltonian (4) of the main text is then diagonal in terms of the $b_{j}$ and $b^{\dagger}_{j}$ fields,
$$\mathcal{H}^{\rm LL}_{0}=\sum_{j=0}^{\infty}\hbar\omega_{j}\left(b^{\dagger}_{j}b_{j}+\frac{1}{2}\right)\;,$$
provided that the modes' wavefunctions $\varphi_{j}(x)$ satisfy the differential equation
\begin{myequation}
-\omega_{j}^{2}\sqrt{v_{s}(x)K(x)}\varphi_{j}(x)=v_{s}(x)K(x)\partial_{x}\left(\frac{v_{s}(x)}{K(x)}\partial_{x}(\sqrt{v_{s}(x)K(x)}\varphi_{j}(x)) \right)\;.
\label{diffeq}
\end{myequation}
This differential equation has the form of a Sturm-Liouville problem, and the functions $\varphi_{j}(x)$ can be chosen real. 
The spatial dependence of the parameters $K$ and $v_s$ can be determined in LDA, assuming a dependence $\mu(n)=\eta n^{\nu}$, yielding $K(x)=K_{0}(1-x^{2}/R^{2})^{1/\nu-1/2}$, and $v_{s}(x)=v_{0}\sqrt{1-x^{2}/R^{2}}$, where $R$ is the Thomas-Fermi radius which is determined from the normalization condition $N=\int_{-R}^{R}\mbox{d}x \,n(x)$, $v_{0}=\sqrt{\nu/2}\omega_{\rm h}R$, and $K_{0}=\sqrt{(\partial\mu/\partial n)|_{TG}/(\partial\mu/\partial n)}|_{x=0}=\sqrt{\hbar^{2}\pi^{2}n(0)^{2-\nu}/m\eta\nu}$. In LDA, we get from the equation of state the following expression for the density: $n(x)=[\eta(\mu-V(x))]^{1/\nu}$, where $V(x)=m\omega_{\rm h}^{2}x^{2}/2$ and $\mu=m\omega_{\rm h}^{2}R^{2}/2$. \\
Eq.~(\ref{diffeq}) has an analytical solution~\cite{citro,abramowitz}:
\begin{myequation}
\begin{split}
 \varphi_{j}(x)&=\sqrt{\frac{j!(j+1/\nu+1/2)}{R\pi\Gamma(j+2/\nu+1)}}2^{1/\nu}\Gamma(1/\nu+1/2)C_{j}^{1/\nu+1/2}(x/R)\;,\\
 (\omega_{j}/\omega_{\rm h})^{2}&=(j+1)(1+j\nu/2)\;,
 \label{phi}
\end{split}
\end{myequation}
where $C_{j}^{a}(x)$ is the Gegenbauer polynomial and $\Gamma(a)$ is the complete $\Gamma$-function. We notice immediately that the lowest eigenvalue $j=0$, that corresponds to the frequency $\omega_0$ of the dipole mode ($j=0$), is given by $\omega_{\rm h}$ for any interaction strength, in agreement with Kohn's theorem.

Let us consider now how the barrier term of the Hamiltonian $\mathcal{H}^{\rm LL}=\mathcal{H}^{\rm LL}_{0}+\mathcal{H}^{\rm LL}_{\rm b}$, $\mathcal{H}^{\rm LL}_{\rm b}=\int_{-\infty}^{\infty} \mbox{d}x\, U_0 \delta(x-d) \rho(x)$, affects the frequency of the dipole mode of $\mathcal{H}^{\rm LL}_{0}$. Considering the LL density field expansion~\cite{cazalilla}, and keeping only the lowest harmonics $l=\pm1$ in $\rho(x)=[n(x)+\partial_{x}\theta(x)/\pi]\sum_{l=-\infty}^{+\infty}e^{2il\theta(x)+2il\pi \int_{-\infty}^{x}{\rm d} x'n(x')}$, we obtain as the most dominant term $\mathcal{H}^{\rm LL}_{\rm{b}}\sim 2 U_{0} n(d)\cos[2\theta(d)+2\pi \int_{-\infty}^{d}\mbox{d}x\,n(x)]$.
Since we are interested only in the dipole excitation mode, if the barrier strength is small compared to the characteristic energy of the dipole mode ($U^{\rm eff}/\ell_{\rm h}<\hbar\omega_{\rm d}$), we can integrate out all the higher modes, $j\geqslant 1$. Considering the Fourier decomposition of the field $\theta (x,t)=\sum_{j=0}^{\infty}e^{-i\omega_{j}t}\theta_{j}(x)$, and taking the zero-temperature average over the vacuum of excitations,  leads to a renormalization of the barrier strength  $U^{\rm eff}=U_{0}\langle 0|\cos(2\sum_{j=1}^{\infty}\theta_{j}(d))|0\rangle$. 
The LL being an effective low-energy field theory, the sum over the modes can not extend up to infinity, but should stop at a certain cut-off $j_{c}$, that is intrinsic of the effective theory and that can not be determined within the theory itself. In our calculations we take it to be proportional to the Fermi energy \footnote{Here we chose the cutoff of the strongly interacting regime; in general $j_c$ is interaction-dependent.}, thus $j_{c}$ is proportional to the number of particles $j_{c}=N/c$, up to some numerical factor $c$ of order $1$. The latter is fixed imposing the matching between the LL solution and the exact TG one in the infinitely strong interaction limit, $K_{0}\rightarrow 1$.
We define $\langle 0|\cos(2\sum_{j=1}^{N/c}\theta_{j}(d))|0\rangle=\exp[-2G(d)]$, where $G(d)=\langle 0| (\sum_{j=1}^{N/c}\theta_{j}(d)))^2 |0\rangle$. Using the mode decompositions~(\ref{modes}), the correlation function takes the form $G(d)=(\pi/2) K(d)v_{s}(d)\sum_{j=1}^{N/c}(1/\omega_{j})|\varphi_{j}(d)|^{2}$. Thus, in general, the effective barrier strength for the lowest mode is given by
\begin{myequation}
U^{\rm eff}(d)=U_{0}e^{-2G(d)}\;.
\end{myequation}
In the particular case $d=0$, when the barrier is placed at the center of the harmonic confinement, and in the $N\gg 1$ limit, the correlation function can be easily evaluated using the analytical solution of Eq.~(\ref{diffeq}), giving $G(0)=\kappa (1/2)\log(N/a)$, where $a$ is a numerical factor proportional to the cut-off parameter $c$.
Using this result, we obtain the expression for the effective barrier strength for the dipole mode:
\begin{myequation}
U^{\rm eff}(0)=U_{0}\left(\frac{a}{N}\right)^\kappa\;,
\end{myequation}
where $\kappa=K(0)v_{s}(0)/\omega_{\rm h}R=K_{0}\sqrt{\frac{\nu}{2}}$.

Having integrated out the higher modes and consequently renormalized the barrier strength, we can rewrite the barrier term of the Hamiltonian in terms only of the field $\theta_{0}$, as $\mathcal{H}^{\rm LL}_{\rm{b}}\sim 2 U^{\rm eff}(d) n(d)\cos[2\theta_{0}(d)+2\pi \int_{-\infty}^{d}\mbox{d}x\,n(x)]$.
Upon a Taylor expansion for small $\theta_{0}$ to second order, substituting Eq.~(\ref{modes}) for the field $\theta_{0}$, and neglecting constant contributions to the Hamiltonian, we get the final, diagonal, expression:
$$\mathcal{H}^{\rm LL}_{\rm{b}}\sim2n(d)U^{\rm eff}(d)\pi v_{s}(d)K(d)\varphi_{0}^{2}(d)\omega_{\rm h}^{-1}\cos(2\pi \int_{-\infty}^{d}\mbox{d}x\,n(x))\left(b_{0}^{\dagger}b_{0} + H.c. \right)\;.$$ This, after substituting the analytical expression of $\varphi_{0}$ given by Eq.~(\ref{phi}), leads us to Eq.~(5) of the main text for the shift of the dipole mode frequency:
\begin{myequation}
\omega_{\rm d}-\omega_{{\rm h}}=\frac{n(d)U^{\rm{eff}}(d)K(d)v_{s}(d)}{\hbar \omega_{\rm h}R}
\!\!\left(\!\frac{(\frac{1}{\nu}\!+\!\frac{1}{2})\Gamma^{2}(\frac{1}{\nu}\!+\!\frac{1}{2})2^{\frac{2}{\nu}+2}}{\Gamma(\frac{2}{\nu}\!+\!1)} \!\right)\cos\left(2\pi \int_{-\infty}^{d}\mbox{d}x\,n(x)\right)\;;
\end{myequation}
which, for $d=0$, simplifies to
\begin{myequation}
\omega_{\rm d}-\omega_{{\rm h}}=(-1)^{N}n(0)U_{0}\left(\frac{a}{N}\right)^{\kappa}K_{0}\sqrt{\frac{\nu}{2}}\frac{1}{\hbar}
\!\!\left(\!\frac{(\frac{1}{\nu}\!+\!\frac{1}{2})\Gamma^{2}(\frac{1}{\nu}\!+\!\frac{1}{2})2^{\frac{2}{\nu}+2}}{\Gamma(\frac{2}{\nu}\!+\!1)} \!\right)\;.
\end{myequation}

\section{IV. Finite-temperature center-of-mass oscillation spectrum}
\label{finiteT}
\subsection{A. Time evolution in the TG limit}
We obtain the oscillation spectrum at finite temperature (Fig 3(d) of the main text) by performing a Fourier analysis of the time evolution of the position of the CM $x_{\text{\tiny{CM}}}(t)=\int {\rm d}x \, x n_{T}(x,t)$, where 
$ n_{T}(x,t)=\sum_{j=0}^{\infty}f(\varepsilon_{j})|\psi_{j}(x,t)|^{2}$. Here,  $f(\varepsilon)$ is the Fermi distribution function at finite temperature for a state of energy $\varepsilon$, $\varepsilon_{j}$ are the single particle energies determined in Section~II, and $\psi_{j}(x,t)$ are the time dependent single particle wavefunctions determined by numerically evolving in real time  with the after-quench Hamiltonian the exact initial-state  wavefunctions determined in Section~II.

\subsection{B. Linear response theory}
The dynamics of the center of mass of the system can also be analyzed, in the limit of small oscillation amplitude, through linear response theory.  This allows to obtain the time evolution of the particle density $n(x,t)=\langle \rho(x,t) \rangle$, where $\rho$ is the density operator in second quantized form, in response to the perturbation operator  $\mathcal{H}_{\rm p}=\int \mbox{d}x V_{\rm p}(x,t)\rho(x)$ where  $V_{\rm p}(x,t)=\Theta(t)\Delta x \partial_x V_{\rm ext}^{t\geqslant 0}(x)$, see Section~I.
Within the framework of linear response theory the corresponding evolution is then given by $\langle\rho(x,t)\rangle=\langle\rho^{I}(x,t)\rangle+\int \mbox{d}x'\int \mbox{d}t' \chi(x,x';t-t') V_{\rm p}(x',t')$, where $\chi(x,x';t,t')=(1/i\hbar)\Theta(t-t')\langle [\rho^{I}(x,t),\rho^{I}(x',t')]\rangle$, and $\rho^{I}(x,t)=e^{i\mathcal{H}^{t\geqslant 0}t/\hbar}\rho(x)e^{-i\mathcal{H}^{t\geqslant 0}t/\hbar}$ gives the unperturbed evolution of the density operator.
Recalling that  the position of the center of mass is given by $x_{\text{\tiny{CM}}}(t)=\int\mbox{d}x\; x\langle \rho(x,t)\rangle$ we readily obtain in Fourier space  $\delta x_{\text{\tiny{CM}}}(\omega)=\int\mbox{d}x\; x\int\mbox{d}x'\; \chi(x,x';\omega)V_{\rm p}(x',\omega)$, with 
\begin{myequation}
 \chi(x,x';\omega)=\frac{1}{\hbar Z(\beta)}\sum_{n\neq m}\langle m|\rho(x)|n\rangle\langle n|\rho(x')|m\rangle e^{-\beta E_{m}}\left(\frac{1}{(\omega-(E_{n}-E_{m})/\hbar))+i0^+}-\frac{1}{(\omega+(E_{n}-E_{m})/\hbar))+i0^+} \right). 
\end{myequation}
Here, $Z(\beta)=\sum_{m}e^{-\beta E_{m}}$, $\beta=1/k_{\rm B}T$, $n$ and $m$ denote many-body states. 

 In the TG limit, using the Bose-Fermi mapping, the density-density response function coincides with the one of a non-interacting Fermi gas: 
\begin{myequation}
\chi(x,x';\omega)=(1/\hbar)\sum_{j\neq k}\psi^{*}_{j}(x)\psi_{k}(x)\psi^{*}_{k}(x')\psi_{j}(x')f(\varepsilon_{j})[1-f(\varepsilon_{k})]\left(\frac{1}{(\omega-(\varepsilon_{k}-\varepsilon_{j})/\hbar))+i0^+}-\frac{1}{(\omega+(\varepsilon_{k}-\varepsilon_{j})/\hbar))+i0^+} \right).
\end{myequation}

\begin{figure}[h]
  \includegraphics[width=0.45\textwidth]{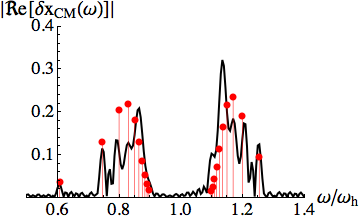}
  \caption{(Color online). Fourier transform of the center of mass oscillation $|\Re\mbox{e}[\delta x_{\text{\tiny{CM}}}(\omega)]|$ as a function of frequency, in the TG limit obtained with the Fourier transform of the real-time evolution (black solid line) and from linear response theory (red vertical lines with dots).}
  \label{peaks}
\end{figure}

In Fig.~\ref{peaks} we show the oscillation spectrum at finite temperature in the TG case obtained with the real-time evolution presented in Section IV.A above and with the linear response theory approach. The results obtained with both methods agree quite well, which is all the more remarkable given the finite simulation time used for the real-time evolution calculation.

\section{Density profiles}

\begin{figure}[h]
  \includegraphics[width=0.5\textwidth]{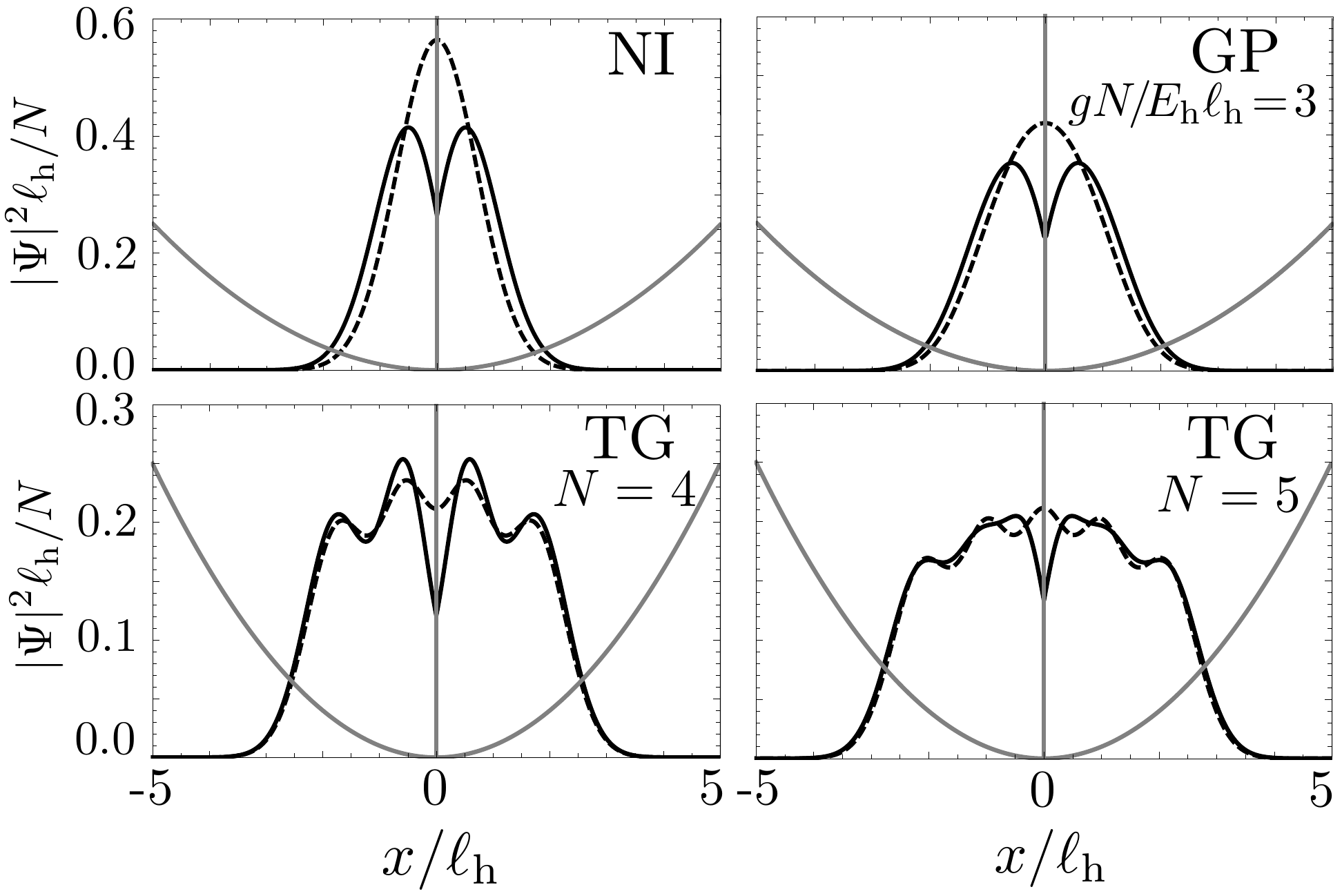}
  \caption{Ground state density for different interparticle interaction regimes and $U_{0}/E_{\rm h}\ell_{\rm h}=0, 1$ (dashed and solid line respectively). In the TG case the barrier creates a notch in the density, in correspondence of a minimum or maximum, depending on whether the number of particles is even or odd. The grey line is a sketch of the potential in dimensionless units and divided by a factor of $50$.}
  \label{density}
\end{figure}

In Fig.~\ref{density} we show the density profiles in the presence and absence of the barrier for different interaction regimes.

\section{Scaling of the dipole frequency shift with number of particles}

\begin{figure}[h]
  \includegraphics[width=0.4\textwidth]{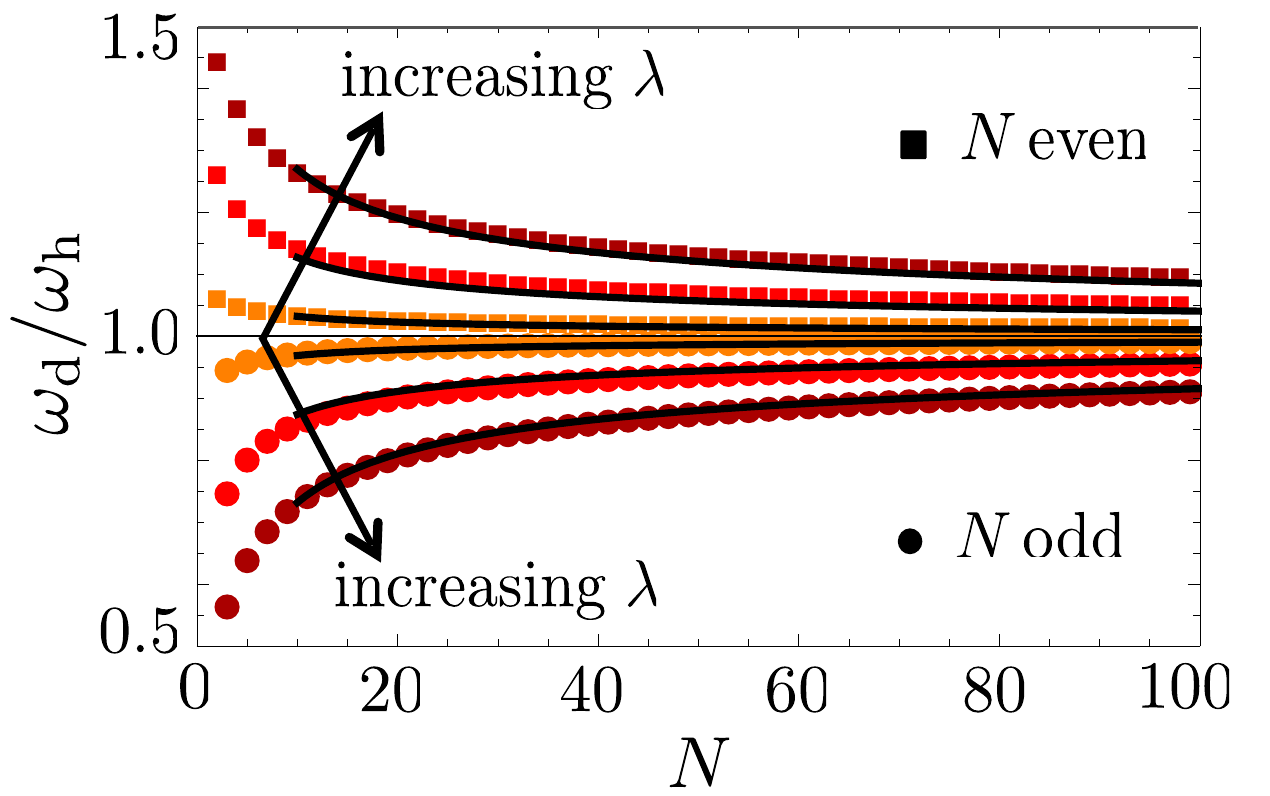}
  \caption{(Color online). Scaling of $\omega_{\rm d}$ as a function of the number of particles $N$ in the TG regime. Different colors correspond to $\lambda=U_{0}/E_{\rm h}\ell_{\rm h}=0.2, 1,$ and $5$ (orange, red, and brown points respectively). For $N$ sufficiently large ($N\gtrsim 10$) the scaling follows the power law $|\omega_{\rm d}-\omega_{\rm h}|\propto 1/\sqrt{N}$ (black solid lines).}
  \label{wvsN}
\end{figure}

In Fig.~\ref{wvsN} we show the scaling behavior of the dipole frequency $\omega_{\rm d}$ as a function of the number of particles $N$, as obtained from the exact calculation in the TG regime. For sufficiently large numbers of particles we find a perfect agreement with the $|\omega_{\rm d}-\omega_{\rm h}|\propto 1/\sqrt{N}$ power law scaling predicted with the LL technique.

\bibliographystyle{aipnum4-1}
\bibliography{biblio_dipolemode}

\end{document}